\documentclass[aps,epsf,preprint,citesort]{revtex4}
\usepackage{latexsym}
\usepackage{amsfonts}
\usepackage{graphicx}

\begin{document}

\begin{center}

{\bf \large Breakdown of the Stokes-Einstein Relation in Supercooled Water}

\end{center}

\bigskip

\begin{center}

Pradeep Kumar$^1$, S. V. Buldyrev$^2$, S. R. Becker,$^3$\\
 P. H. Poole$^4$, F. W. Starr$^3$, and H. E. Stanley$^1$

\bigskip

\noindent{\scriptsize $^{1}$Center for Polymer Studies and Department
of Physics, Boston University,~Boston, MA 02215 USA \\ $^{2}$ Yeshiva
University, Department of Physics, 500 West 185th Street, New York,
New York 10033, USA\\ $^3$Department of Physics, Wesleyan University,
Middletown, CT 06459, USA\\ $^4$ Department of Physics, St. Francis
Xavier University, Antigonish, Nova Scotia B2G 2W5, Canada\\}
%Jawaharlal Nehru Center for Advanced Scientific Research, Jakkur Campus,
%Bangalore, India\\}

\end{center}

\noindent{\scriptsize PACS:} 05.40.-a \hfill {\scriptsize REVISED:} 09 Jan

\bigskip

\begin{center}

ABSTRACT

\end{center}

\bigskip

\noindent{\bf Supercooled water exhibits a breakdown of the
 Stokes-Einstein relation between the diffusion constant $D$ and the
 alpha relaxation time $\tau_{\alpha}$. For water simulated with the
 TIP5P and ST2 potentials, we find that the temperature of the
 decoupling of diffusion and alpha relaxation correlates with the
 temperature of the maximum in specific heat that corresponds to
 crossing the Widom line $T_W(P)$.  Specifically, we find that our
 results for $D\tau_{\alpha}/T$ collapse onto a single master curve if
 temperature is replaced by $T-T_W(P)$, where $T_W(P)$ is the
 temperature where the constant-pressure specific heat achieves a
 maximum. Also, we find agreement between our ST2 simulations and
 experimental values of $D\tau_{\alpha}/T$. We further find that the
 size of the mobile molecule clusters (dynamical heterogeneities)
 increases sharply near $T_W(P)$. Moreover, our calculations of mobile
 particle cluster size $\langle n(t^*)\rangle_w$ for different
 pressures, where $t^*$ is the time for which the mobile particle
 cluster size is largest, also collapse onto a single master curve if
 $T$ is replaced by $T-T_W(P)$. The crossover to a more locally
 structured low density liquid (LDL) environment as $T\rightarrow
 T_W(P)$ appears to be well correlated with both the breakdown of the
 Stokes-Einstein relation and the growth of dynamic heterogeneities.}

\bigskip

A 17th century study of the density maximum at $4^{\circ}$C~\cite{tmd}
demonstrates the long history of water science
\cite{mishima98,pgdbook,pabloreview,angell-review}.  Since that time,
dozens of additional anomalies of water have been
discovered~\cite{chaplin}, including the sharp increase upon cooling
of both the constant-pressure specific heat $C_P$ and the isothermal
compressibility $K_T$.  These anomalies of water become more
pronounced as water is supercooled. To explain these properties, a
liquid-liquid (LL) critical point has been proposed \cite{poole1}.
Emanating from this critical point there must be loci of extrema of
thermodynamic response functions such as $C_P$ and $K_T$.  These loci
must coincide as the critical point is approached, since response
functions are proportional to powers of the correlation length, and
the locus of the correlation length maxima asymptotically close to the
critical point defines the Widom line.

In the supercooled region of the pressure-temperature phase diagram,
the dynamic properties of water show dramatic
changes~\cite{chenmallamace2006,prielmeierXX}. One basic relation
among dynamic properties is the Stokes-Einstein (SE) relation
\begin{equation}
D=\frac{k_BT}{6\pi\eta a},
\label{eq:stoke}
\end{equation}
where $D$ is the diffusion constant, $T$ is the temperature, $k_B$ is
the Boltzmann constant, $\eta$ is the viscosity and $a$ is the
effective hydrodynamic radius of a molecule. This expression provides
a relation between mass and momentum transport of a spherical object
in a viscous medium.  The SE relation describes nearly all fluids at
$T\gtrsim 1.2-1.6\, T_g$, where $T_g$ is the glass transition
temperature, and since the hydrodynamic radius $a$ is roughly
constant, $D\eta/T$ is approximately independent of $T$
\cite{egelstaff,se-obedience1,se-obedience2}. However, in most
liquids, for $T\lesssim 1.2-1.6\,T_g$ $D\eta/T$ is no longer a
constant
\cite{sillescu-rev,OTP,Cicerone95,berthier,pollack,Pollack85,ediger-rev,sillescu-rev,fujara,ediger,sillescu}.
For the case of water, the breakdown of the SE relation occurs about
$1.8 T_g$, in the same temperature region in which many of the unusual
thermodynamic features of water
occur~\cite{angell-review,chenmallamace2006,sas}.

\begin{figure}
\begin{center}
\includegraphics[width=10cm]{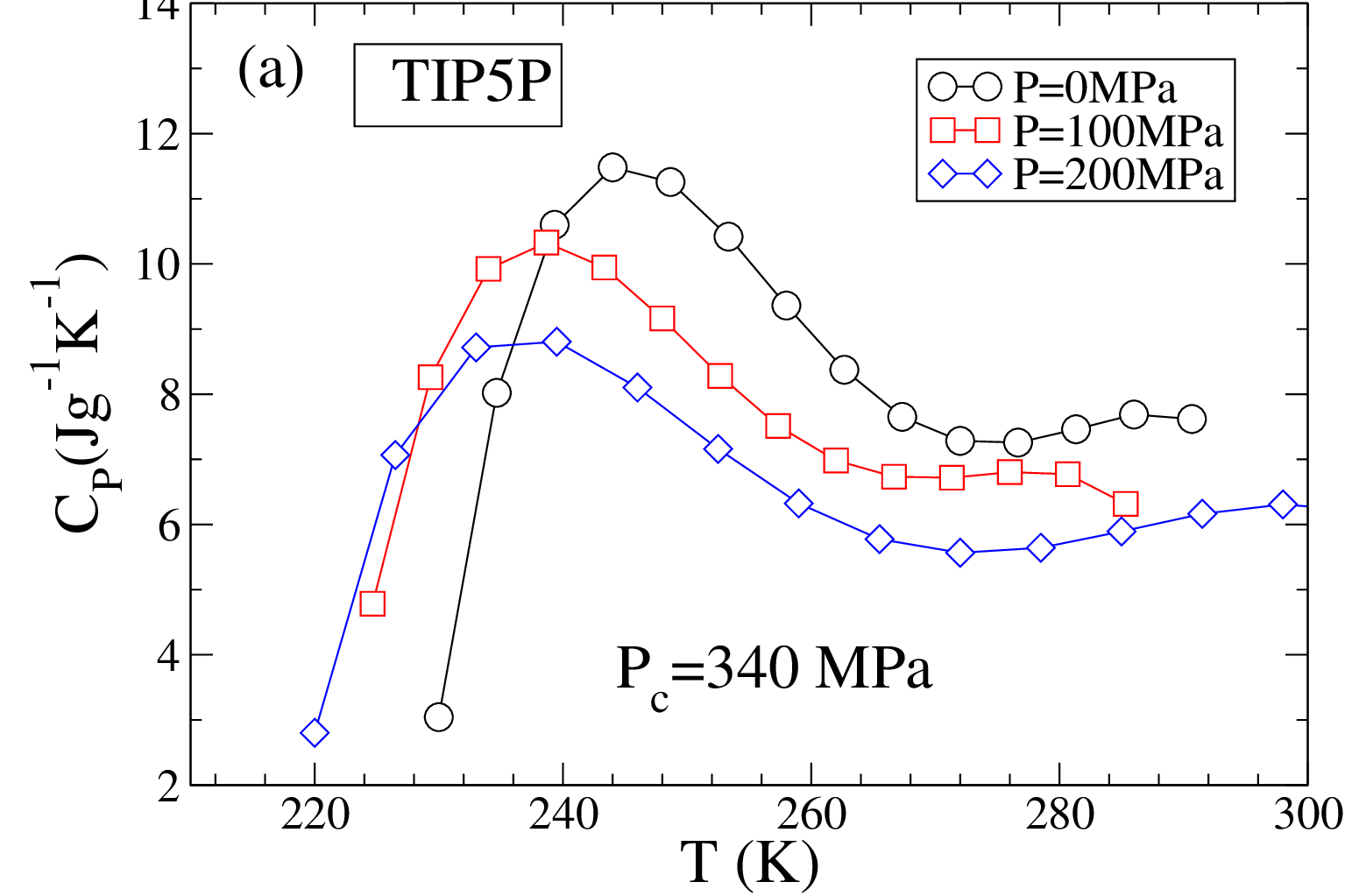}
\includegraphics[width=10cm]{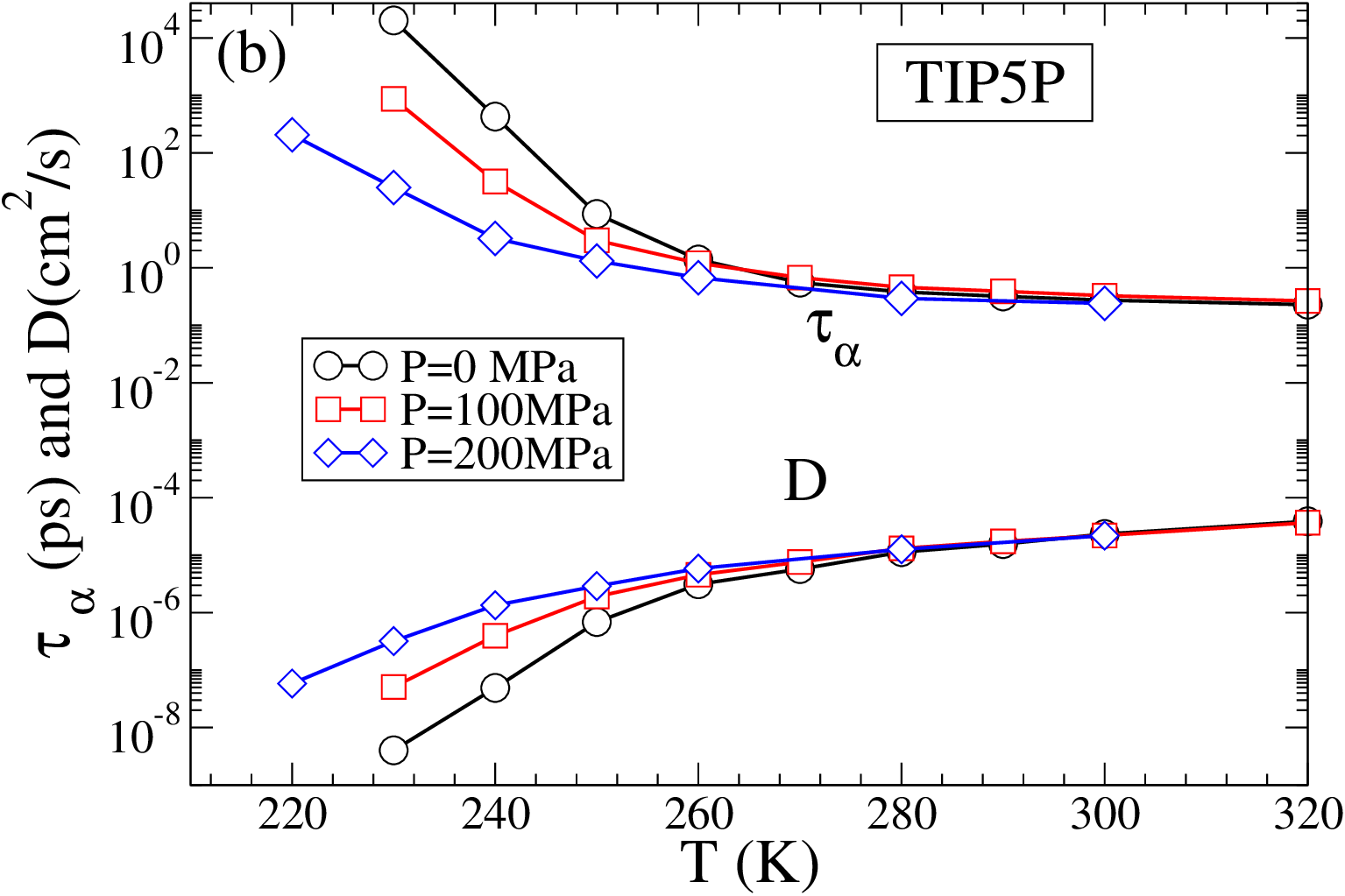}
\end{center}
\caption{Temperature dependence of (a) $C_P$ and (b) $\tau_{\alpha}$
 and $D$ for $P=0$~MPa, 100~MPa and 200~MPa. We note that while the
 temperature of the maximum decreases with pressure, the value of
 $C_P(T)$ at the maximum also decreases. This could be an artifact of
 the TIP5P potential which is known not to correctly reproduce the
 radial distribution function at high pressures.}
\label{fig:dtau}
\end{figure}

Our aim is to evaluate to what degree the SE breakdown can be
correlated with the presence of thermodynamic anomalies and the onset
of spatially heterogeneous dynamics, and how these features relate to
the location of the Widom
line~\cite{stillinger,tarjus,ngai,berthier,weeks2000}. From prior
studies of water, we can already form an expectation for the
correlation between the SE breakdown and the Widom line by combining
three elements: (i) the Widom line is approximately known from the
extrapolated power-law divergence of $K_T$~\cite{kano-angell}; (ii)
the locus of points $T_D(P)$ where $D$ extrapolates to zero is also
known, and nearly coincides with $T_W(P)$ at low pressures (see Fig.1
of Ref.~\cite{sss}); (iii) the SE relation has been found to fail in
liquids generally at the temperature $T_D(P)$~\cite{rossler}.
Combining these three results, one might not be surprised if the
breakdown of the SE relation should occur near to the Widom line for
$P<P_C $, and it should continue to follow $T_D(P)$ for $P>P_C$.  We
will see that our results are consistent with this expectation, but
reveal some unexpected insights. To address these issues, we perform
molecular dynamics (MD) simulations of $N=512$ waterlike molecules
interacting via the TIP5P potential \cite{JorgensenXX,YamadaXX}, which
exhibits a LL critical point at approximately $T_C\approx217$~K and
$P_C\approx340$~MPa \cite{YamadaXX,Paschek05}. We carry out
simulations in the $NPT$ ensemble for three different pressures
$P=0$~MPa, $100$~MPa, and $200$~MPa, and for temperatures $T$ ranging
from $320$~K down to $230$~K for $P=0$ and $100$~MPa, and down to
$220$~K for $200$~MPa.  We also analyze MD simulations of $N=1728$
waterlike molecules interacting via the ST2 potential~\cite{st2,bps},
which displays a LL critical point at $T_C\approx245$~K and
$P=180$~MPa~\cite{poole2}.  The simulations for the ST2 model are
performed in the $NVT$ ensemble~\cite{bps}. To locate the Widom line, we
first analyze the isobaric heat capacity $C_P$ for the TIP5P model
[Fig.~\ref{fig:dtau}(a)]. For the ST2 model, we use the results for
the Widom line from Ref.~\cite{poole2}.

We next explore the possible relation between the Widom line and the
breakdown of the SE relation(Eq.~\ref{eq:stoke}). We first calculate the
diffusion constant via its asymptotic relation to the mean-squared
displacement,
\begin{equation}
D\equiv\lim_{t\to \infty}{\langle |{\bf r}_j(t)-{\bf r}_j(0)|^2
\rangle\over 6t},
\label{eq:diff}
\end{equation}
where ${\bf r}_j(t)$ is the position of oxygen $j$ at time $t$.  It is
difficult to accurately calculate the viscosity $\eta$ in simulations,
so we instead calculate the alpha relaxation time $\tau_{\alpha}$
(which is expected to have nearly the same $T$ dependence as $\eta$)
as the time at which the coherent intermediate scattering function
\begin{equation}
F(q,t) \equiv \frac{\langle\rho({\bf q},t)\rho(-{\bf
q},0)\rangle}{\langle\rho({\bf q},0)\rho(-{\bf q},0)\rangle},
\end{equation}
decays by a factor of $e$. Here $\rho({\bf
 q},t)\equiv\sum_j^N\exp[-i{\bf q} \cdot {\bf r}_j(t)]$ is the Fourier
 transform of the density at time $t$, ${\bf r}_j(t)$ is the position
 of oxygen $j$ at time $t$, ${\bf q}$ is a wave vector and the
 brackets denote an average over all $|{\bf q}|=q$ and many
 equilibrium starting configurations. We calculate $F(q,t)$ at the
 value of $q$ of the first maximum in the static structure factor
 $F(q,0)$. It is important that we use the coherent scattering
 function (as opposed to the incoherent, or self-scattering function),
 since we want to capture collective relaxation that contributes to
 $\eta$. Hence the SE relation can be written as
\begin{equation}
{D\tau_\alpha\over T}={\rm constant}.
\label{eq:se}
\end{equation}
\noindent
We see that both $\tau_{\alpha}$ and $D$ (Fig.~\ref{fig:dtau}(b))
display rapid changes at low $T$.

\begin{figure}
\begin{center}
\includegraphics[width=7cm]{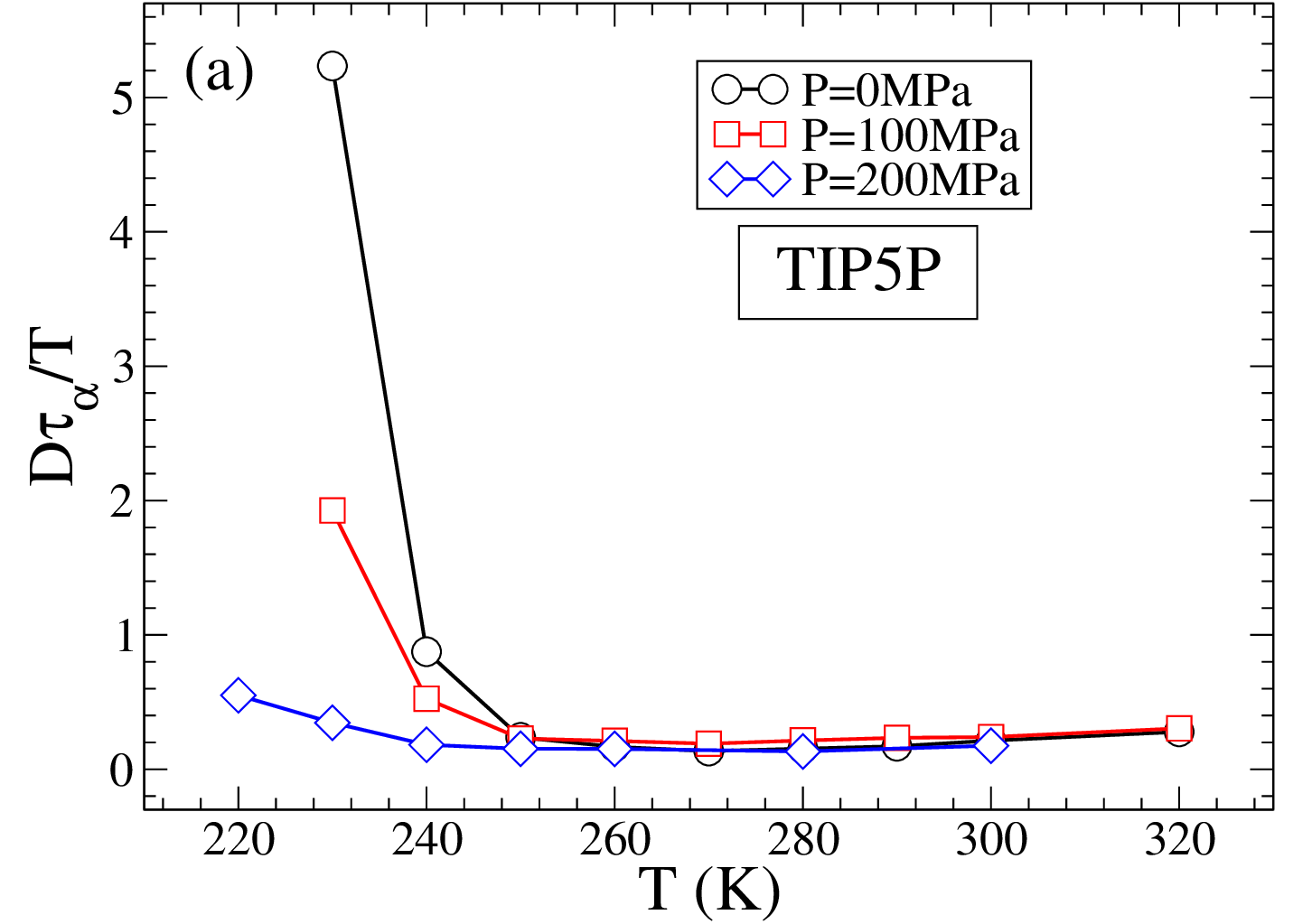}
\includegraphics[width=7cm]{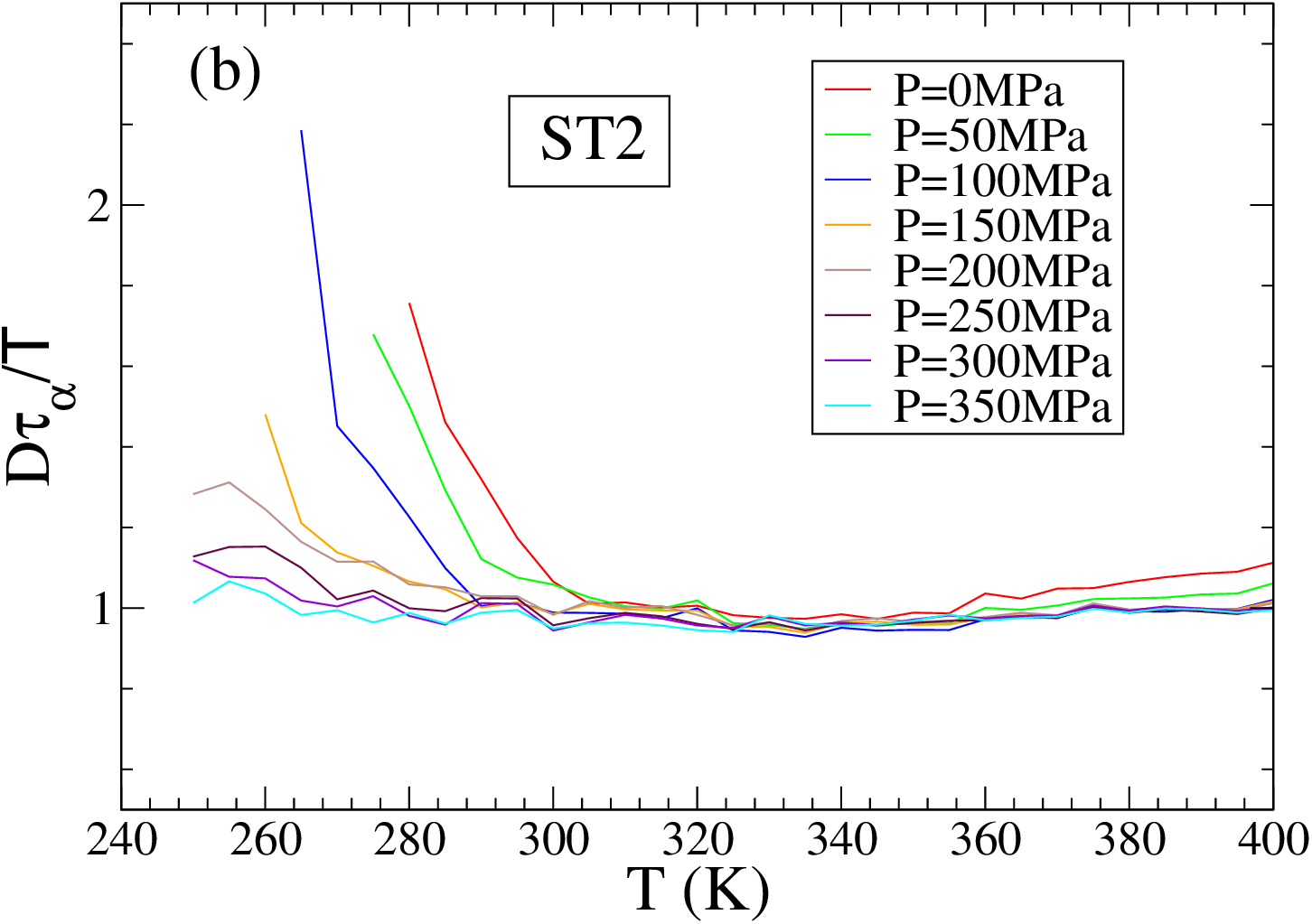}
\includegraphics[width=7cm]{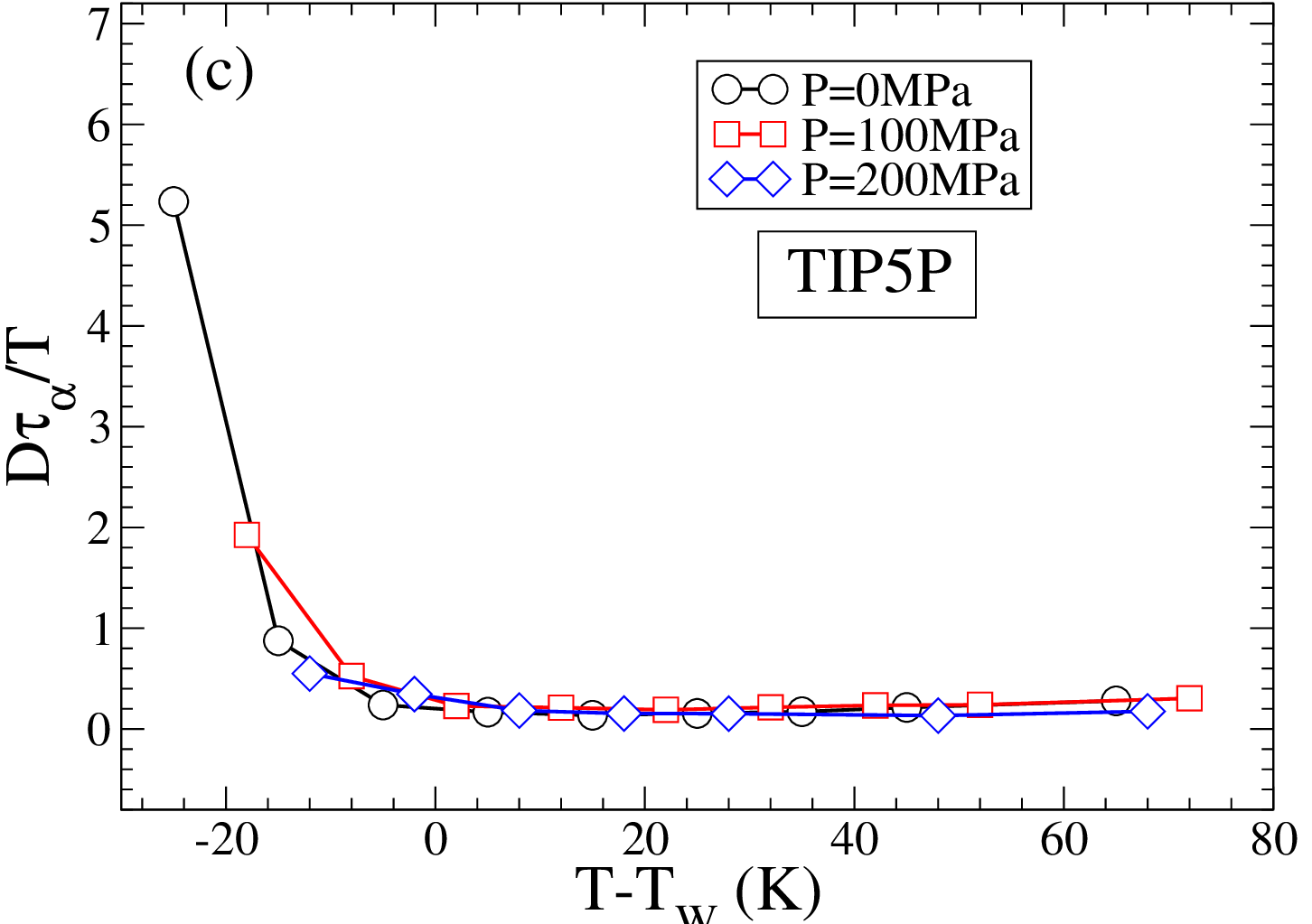}
\includegraphics[width=7cm]{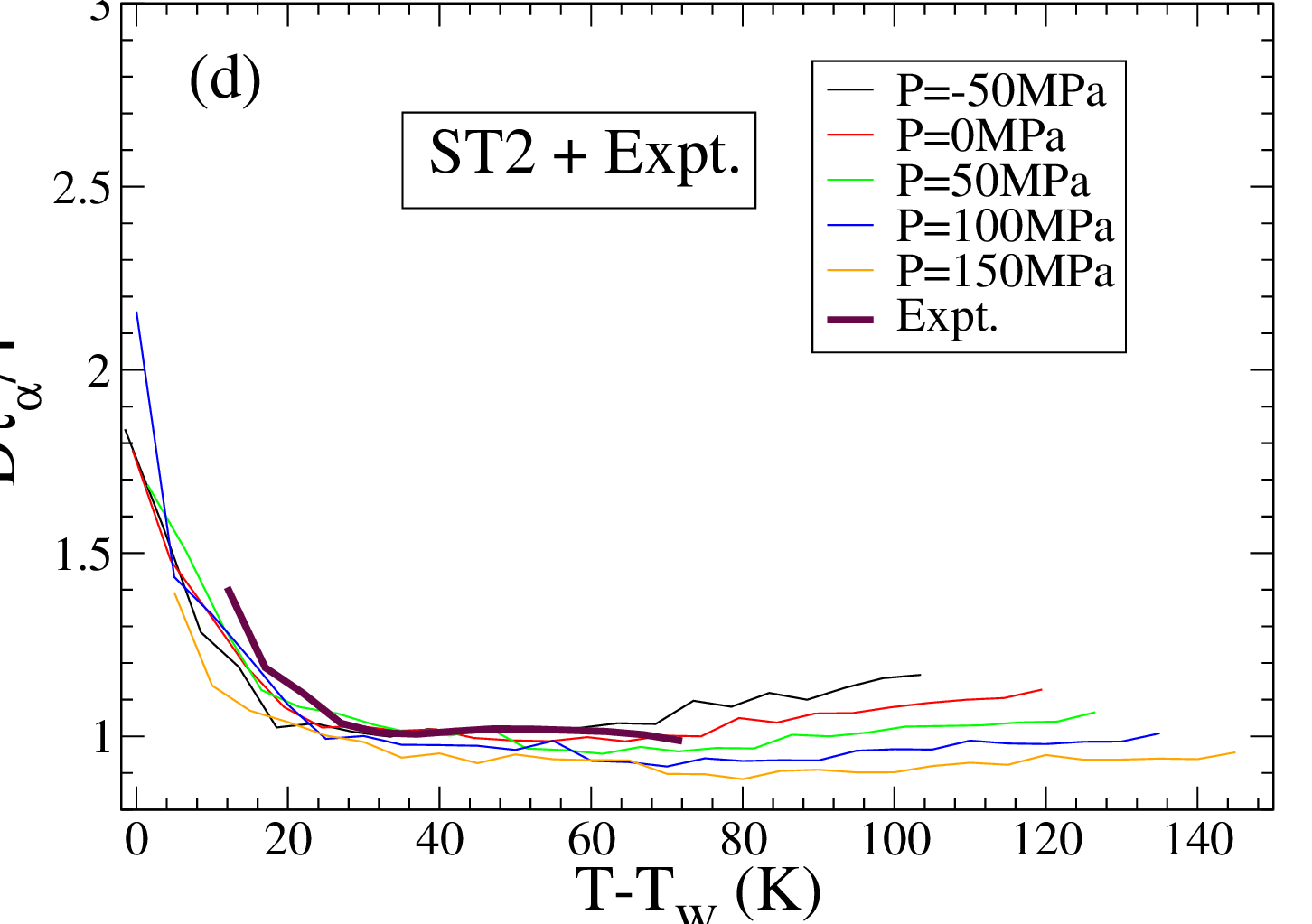}
\end{center}
\caption{(a) $D\tau_{\alpha}/T$ as a function of $T$ for $P=0$~MPa,
 $100$~MPa and $200$~MPa for the TIP5P model. For all panels,
 $D\tau_{\alpha}/T$ is scaled by its high $T$ value to facilitate
 comparison of the different systems. (b) Analog of
 Fig.~\ref{fig:dtau-rescale}(a) for the ST2 model. (c)
 $D\tau_{\alpha}/T$ as a function of $T-T_W(P)$ for TIP5P. (d) Analog
 of Fig.~\ref{fig:dtau-rescale} (c) for the ST2 model including the
 experimental curve of $D\eta/T$ for water.}
\label{fig:dtau-rescale}
\end{figure}

Figures~\ref{fig:dtau-rescale}(a) and \ref{fig:dtau-rescale}(b) show
$D\tau_{\alpha}/T$ as a function of $T$. At high $T$,
$D\tau_{\alpha}/T$ remains approximately constant. At low $T$,
$D\tau_{\alpha}/T$ increases indicating a breakdown of the SE relation
(\ref{eq:stoke}), which occurs in the same $T$ region where the
thermodynamic anomalies develop, and near the Widom line $T_W(P)$.  To
test if there is a correlation between the SE breakdown and $T_W(P)$,
we plot $D\tau_{\alpha}/T$ against $T-T_W(P)$
[Figs.~\ref{fig:dtau-rescale}(c) and (d)] and find the unexpected
result that all the curves for different pressures overlap within the
limits of our accuracy for both TIP5P and ST2 potentials.  Hence
$D\tau_{\alpha}$ is a function only of $T-T_W(P)$, from which it
follows that the locus of the temperature of the breakdown of the SE
relation is correlated with $T_W(P)$.

\begin{figure}
\begin{center}
\includegraphics[width=10cm]{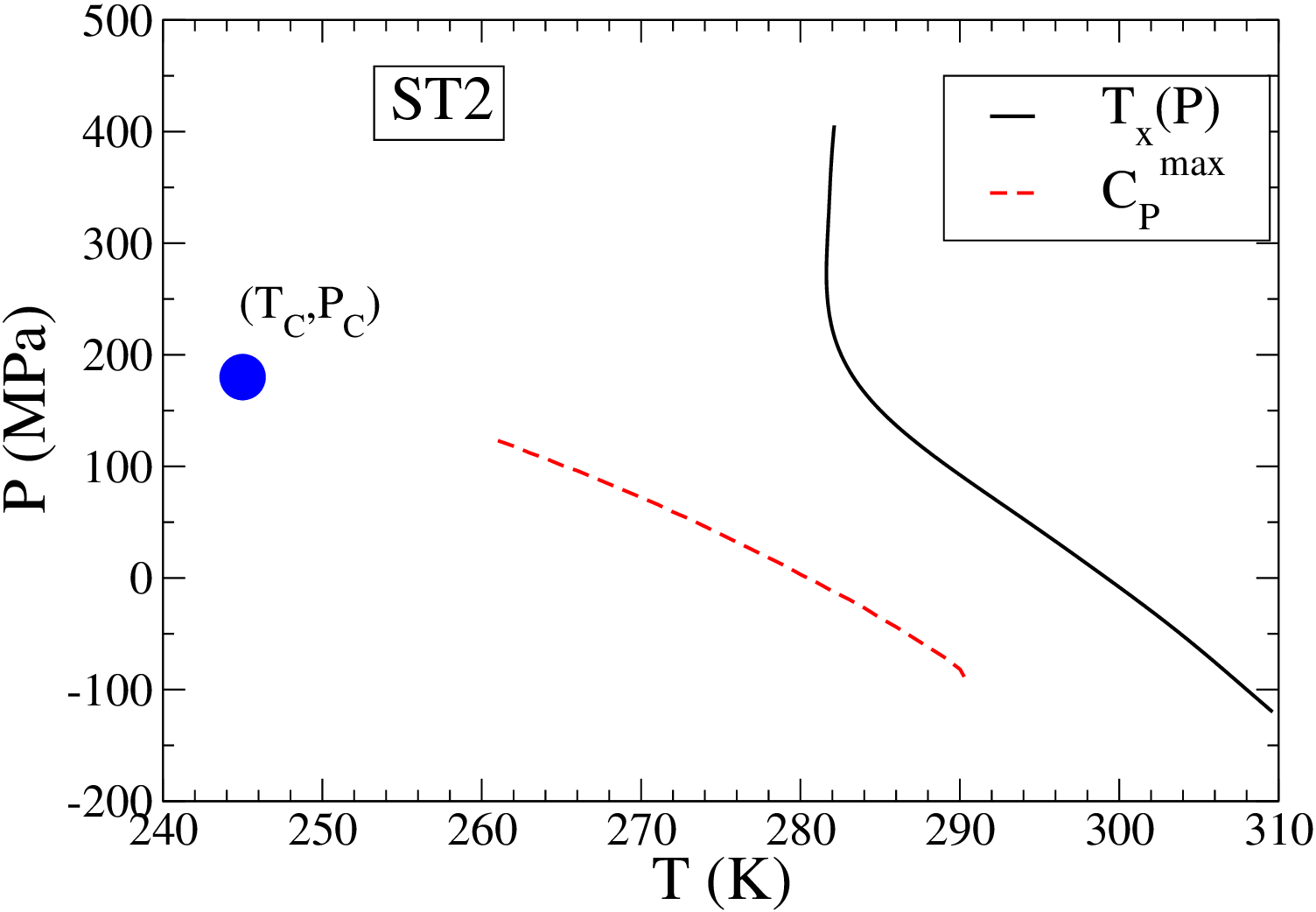}
\end{center}
\caption{Locus in P-T plane of $C_P^{max}$ and $T_{\times}(P)$ for the
ST2 model. Filled circle denotes the liquid-liquid critical point
$(T_C,P_C)$ in ST2 model.}
\label{fig:loci}
\end{figure}

To better quantify the temperature where the SE relation breaks down,
we use the fact that when the SE relation fails it can be replaced by
a ``fractional'' SE relation $D(\tau_{\alpha}/T)^{\xi}={\rm
const}$~\cite{sillescu-rev,bps}. By plotting parametrically $\log(D)$
as a function of $\log(\tau_{\alpha}/T)$, one can identify the
crossover temperature $T_{\times}(P)$ between the two regions by the
intersection of the high $T$ SE behavior and the low $T$ fractional SE
behavior [Fig.~\ref{fig:loci}]. We confirm that the same collapse of
$D\tau_{\alpha}/T$ can be found by replacing $T$ with
$T-T_{\times}(P)$, demonstrating (Fig.~\ref{fig:loci}) that the loci
of SE breakdown defined by $T_{\times}(P)$ and $T_W(P)$ track each
other for $P<P_C$. There is also some difference between the ST2 and
TIP5P models in the relative location of the breakdown of the SE
relation and $T_W(P)$, evidenced by the fact that the magnitude of the
SE breakdown is different at $T=T_W(P)$. As a result, the data for the
two models will not collapse when plotted together. Moreover, we find
that $D\eta/T$ from the experimental data plotted against $T-T_W(P)$
[Fig.~\ref{fig:dtau-rescale}(d)], coincides with the ST2 results,
suggesting that the data collapse should exist for real water when $T$
is replaced by $T-T_W(P)$~\cite{footnote,angell-review81}.

The above analysis has primarily focused on the behavior for $P<P_C$,
where there is a Widom line.  For $P>P_C$, water behaves similarly to
simple glass forming liquids, and we expect the breakdown of the SE
relation for water to be similar to other simple liquids.
Specifically, the SE relation should break down at $T\approx 1.2-1.6
T_g$~\cite{sillescu-rev}. This breakdown approximately coincides with
the temperature where the mode coupling description of the dynamics
fails ~\cite{rossler,sillescu-rev}.  For the SPC/E model of water, the
mode coupling temperature ($T_{MCT}(P)$) locus has been evaluated, and
the slope in the $P$-$T$ plane of this locus changes from negative to
positive for $P>P_C$~\cite{fwstarr99} (the slope is positive for
simple liquids).  Correspondingly, there is also a decoupling of $D$
and $\tau_\alpha$ near $T_{MCT}(P)$~\cite{fwstarr99}. We find the same
behavior for the ST2 model.

Since we find a correlation between $T_W(P)$ and the breakdown of the
SE relation, the hypothesized connection between the SE breakdown and
the onset of dynamical heterogeneities (DH) suggests a connection
between $T_W(P)$ and the onset of DH. To investigate the behavior of
the dynamic heterogeneities, we study the clusters formed by the $7\%$
most mobile molecules~\cite{donati}, defined as molecules with the
largest displacements during a certain interval of time of length
$t$. The clusters of the most mobile molecules are defined as
follows. If in a pair of the most mobile molecules determined in the
interval $[t_0,t_0+t]$, two oxygens at time $t_0$ are separated by
less than the distance corresponding to position of the first minimum
in the pair correlation function ($0.315$~nm in TIP5P and $0.350$~nm
in ST2), this pair belongs to the same cluster.

\begin{figure}
\begin{center}
\includegraphics[width=7cm]{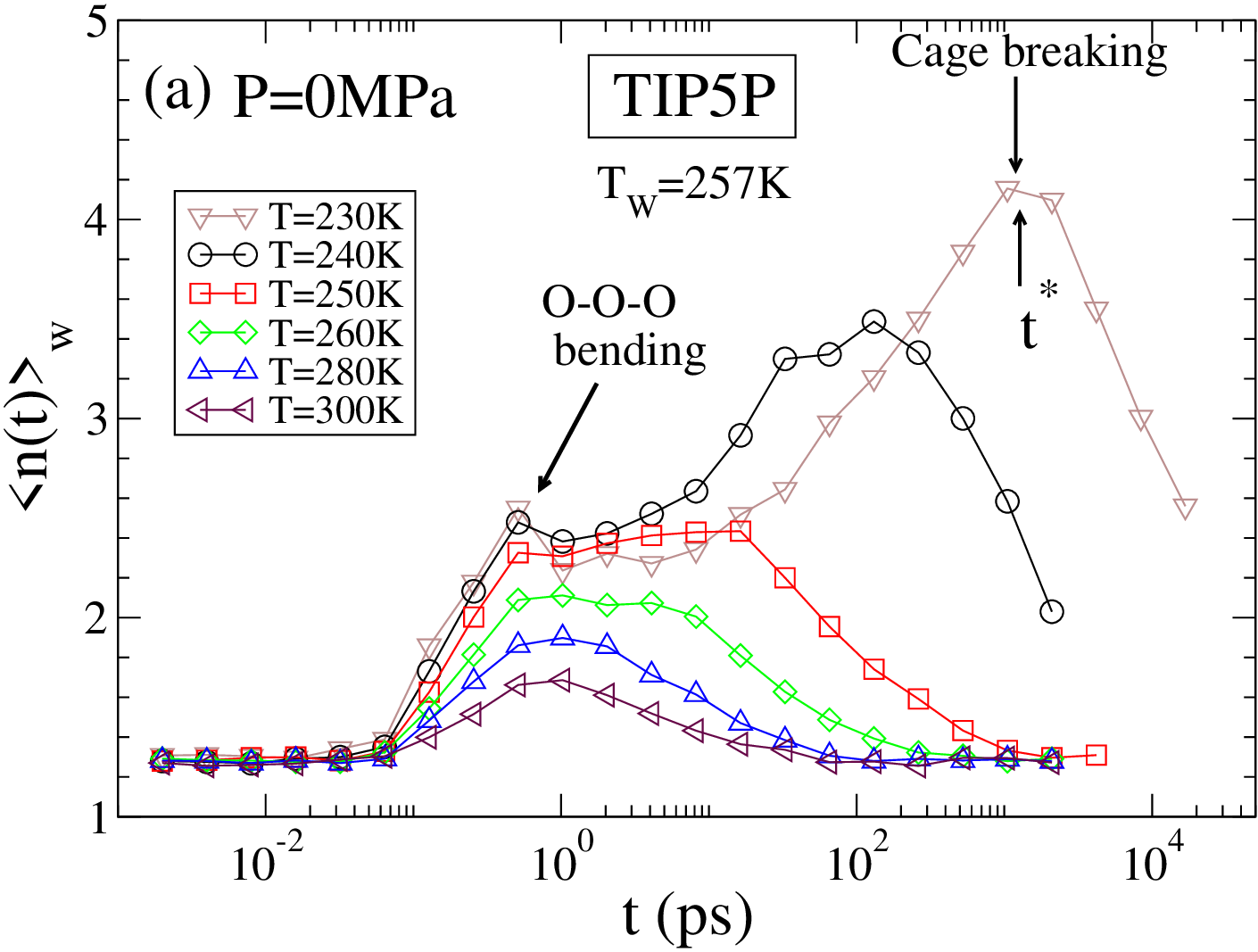}
\includegraphics[width=7cm]{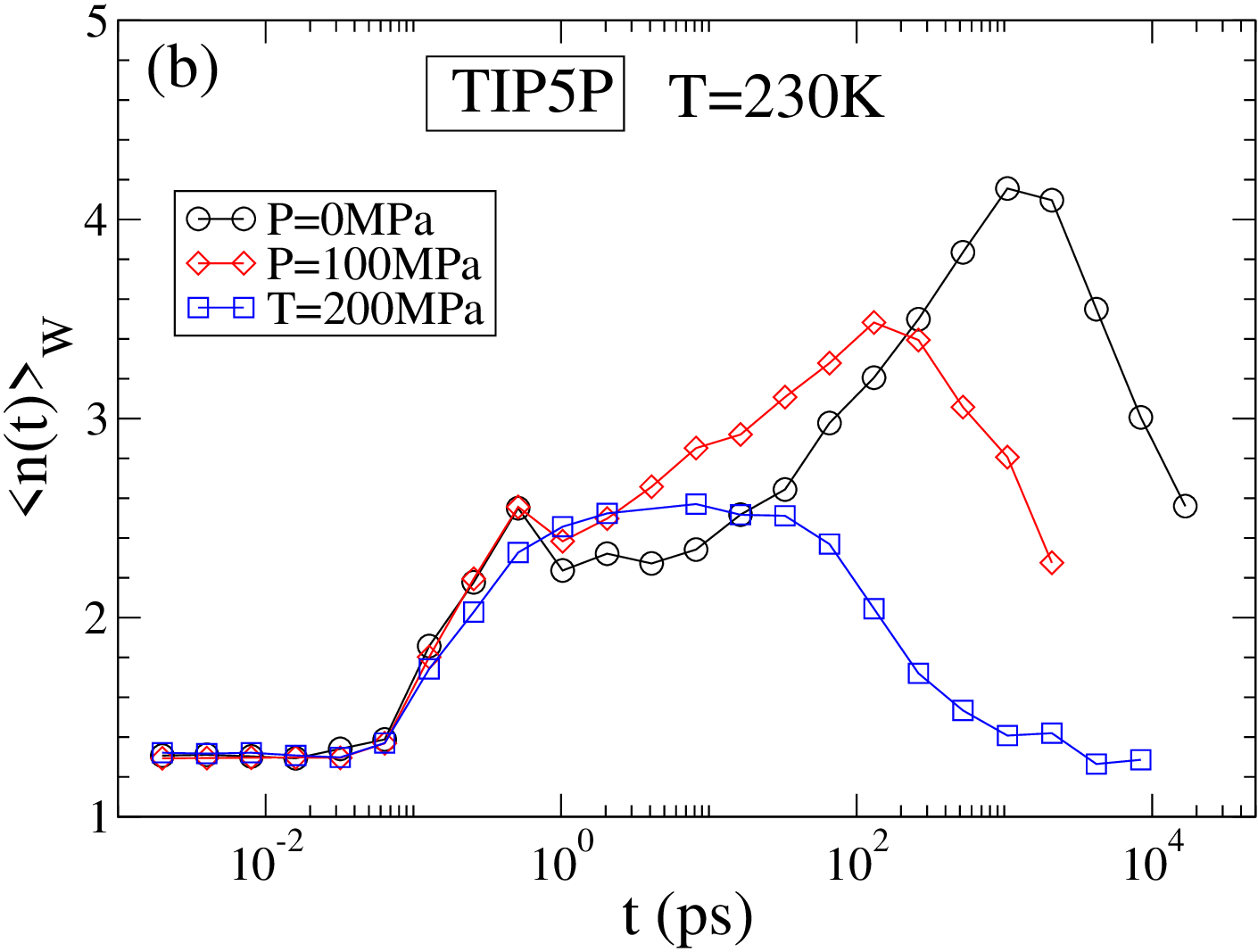}
\includegraphics[width=7cm]{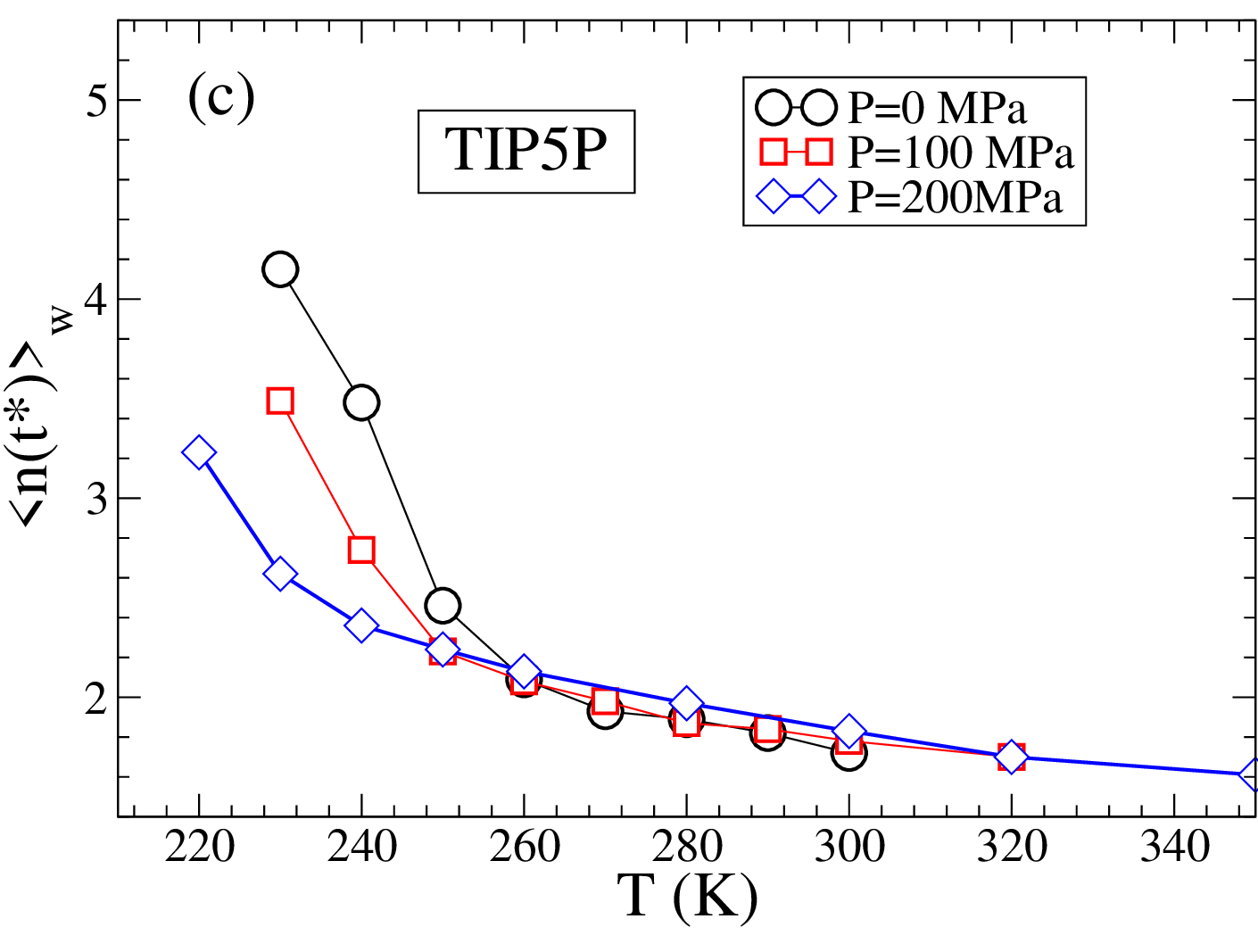}
\includegraphics[width=7cm]{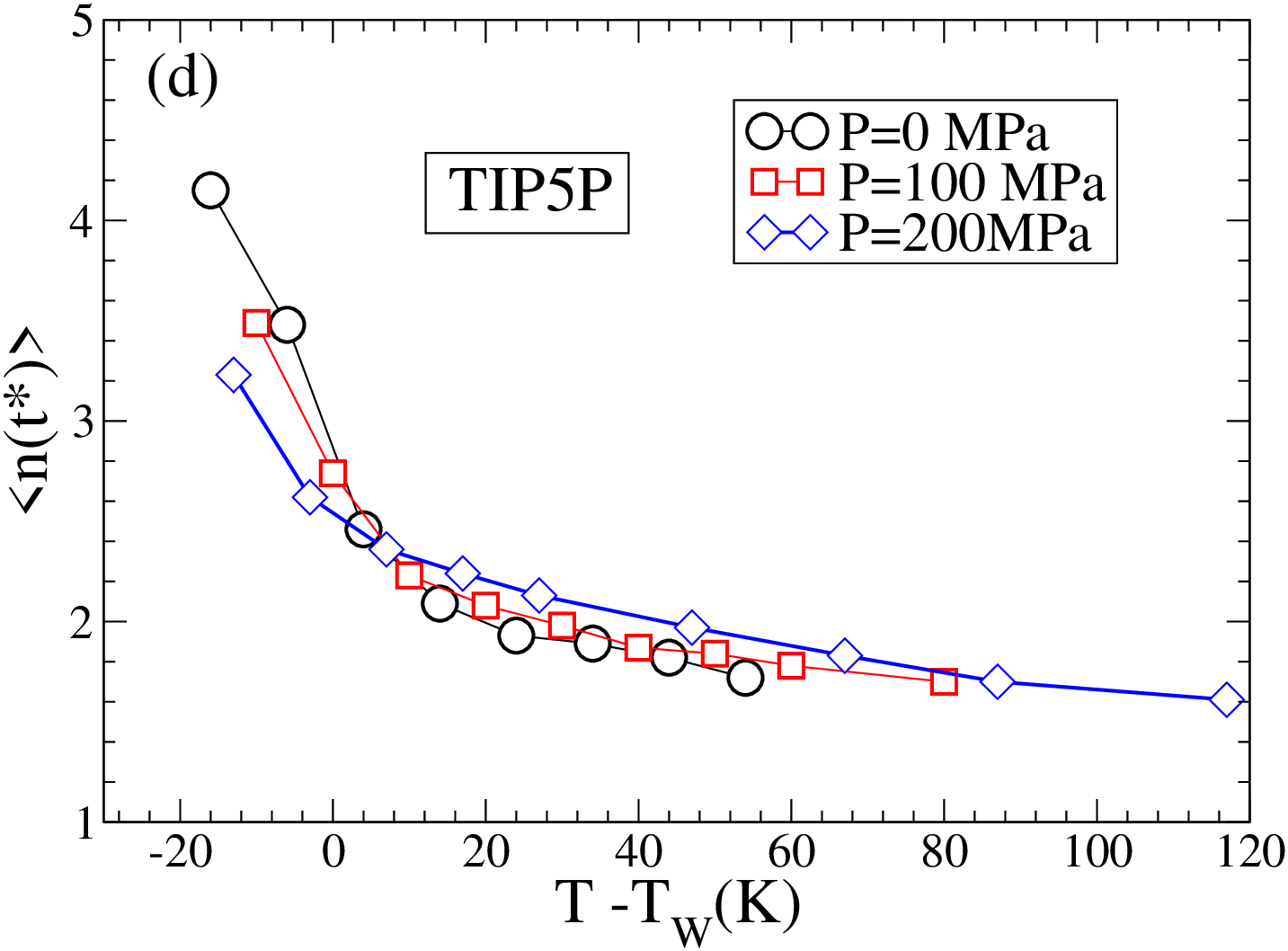}
\end{center}
\caption{Mobile particle cluster size $\langle n(t)\rangle_w$ (a) for
different $T$ at $P=0$~MPa and (b) for different $P$ at $T=230$~K.
(c) $\langle n(t^*)\rangle_w$ for $P=0$~MPa,~$100$~MPa and
$200$~MPa. (d) $\langle n(t^*)\rangle$ as a function of $T-T_W(P)$ for
three different $P$. All these plots are for the TIP5P model.}
\label{fig:clus}
\end{figure}

The weight averaged mean cluster size
\begin{equation}
\langle n(t)\rangle_w \equiv \frac{\langle n^2(t)\rangle}{\langle
n(t)\rangle},
\end{equation}
measures the cluster size to which a randomly chosen molecule belongs,
where $\langle n(t)\rangle$ is the number averaged mean cluster
size. We show $\langle n(t)\rangle_w$ in Fig.~\ref{fig:clus} for TIP5P
as a function of observation time interval $t$ for different $T$ at
$P=0$~MPa [Fig.~\ref{fig:clus}(a)]. The behavior at higher $P$ is
qualitatively the same [Fig.~\ref{fig:clus}(b)]. At low $T$, $\langle
n(t)\rangle_w$ has a maximum at the time $t^*$ associated with the
breaking of the cage formed by the neighboring molecules (see
\cite{gbss} and the references therein).  Both the magnitude and the
time scale $t^*$ of the peak grow as $T$ decreases.  At high $T$, this
peak merges and becomes indistinguishable from a second peak with
fixed characteristic time $\approx 0.5$~ps.  By evaluating the
vibrational density of states, we associate this feature with a low
frequency vibrational motion of the system, probably the O-O-O bending
mode~\cite{tsai2005}.

To probe the temperature dependence of $\langle n(t) \rangle_w$, we plot the peak
value $\langle n(t^*)\rangle_w$ in Fig.~\ref{fig:clus}(c) as a
function of $T$ for $P=0$~MPa, 100~MPa, and 200~MPa for the TIP5P
model. At high $T$, $\langle n(t^*)\rangle_w$ is nearly constant,
since at high $T$, clusters result simply from random motion of the
molecules.  Upon cooling near the Widom line, $\langle
n(t^*)\rangle_w$ increases sharply.  When $\langle n(t^*)\rangle_w$ is
plotted as a function of $T-T_W(P)$ [see Fig.~\ref{fig:clus}(d)], we
find that (similarly to the behavior of $D\tau_{\alpha}/T$) the three
curves for $P=0$~MPa, $P=100$~MPa, and $P=200$~MPa overlap, and it is
apparent that the pronounced increase in $\langle n(t^*)\rangle_w$
occurs for $T\approx T_W(P)$.

\begin{figure}
\begin{center}
\includegraphics[width=10cm]{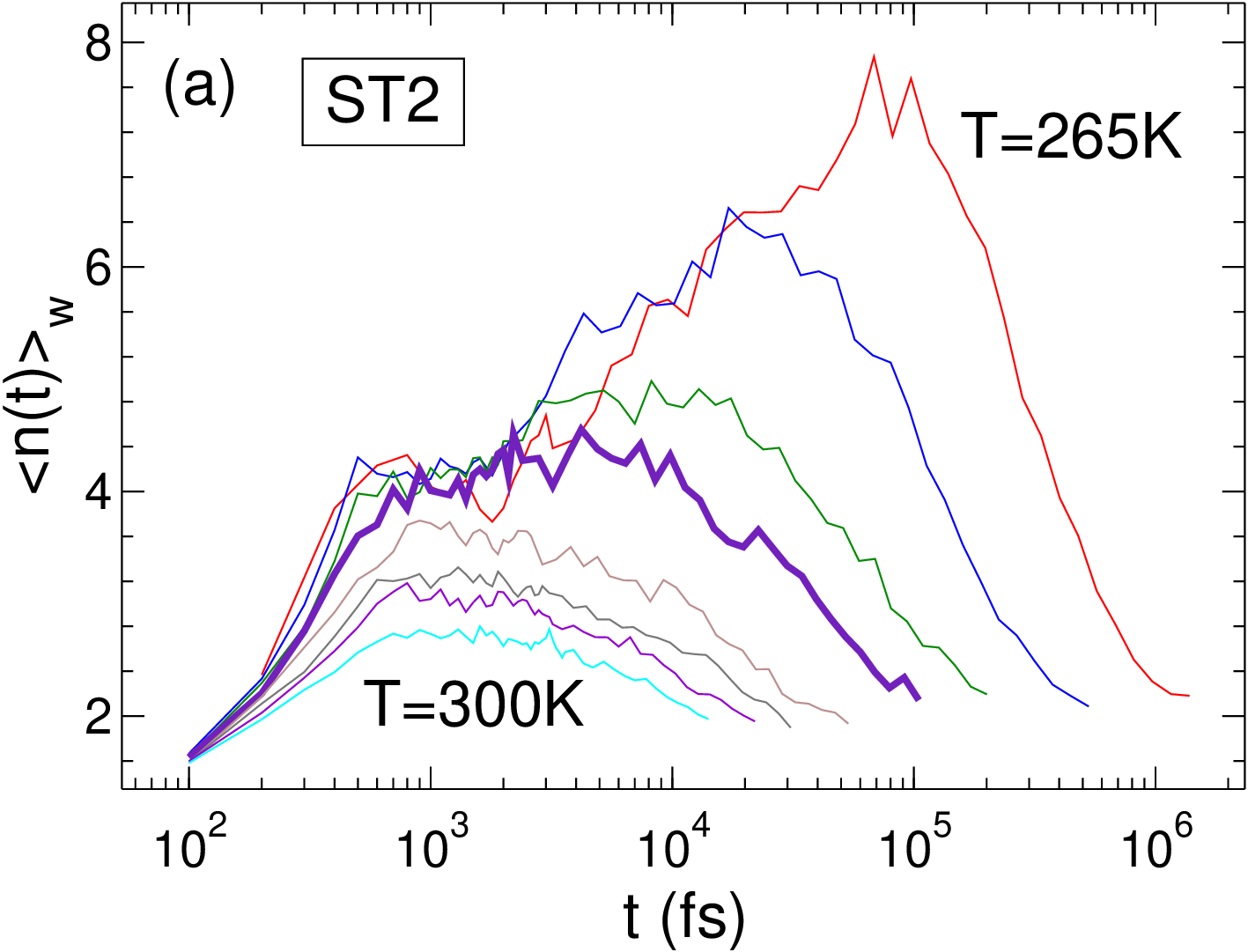}
\includegraphics[width=10cm]{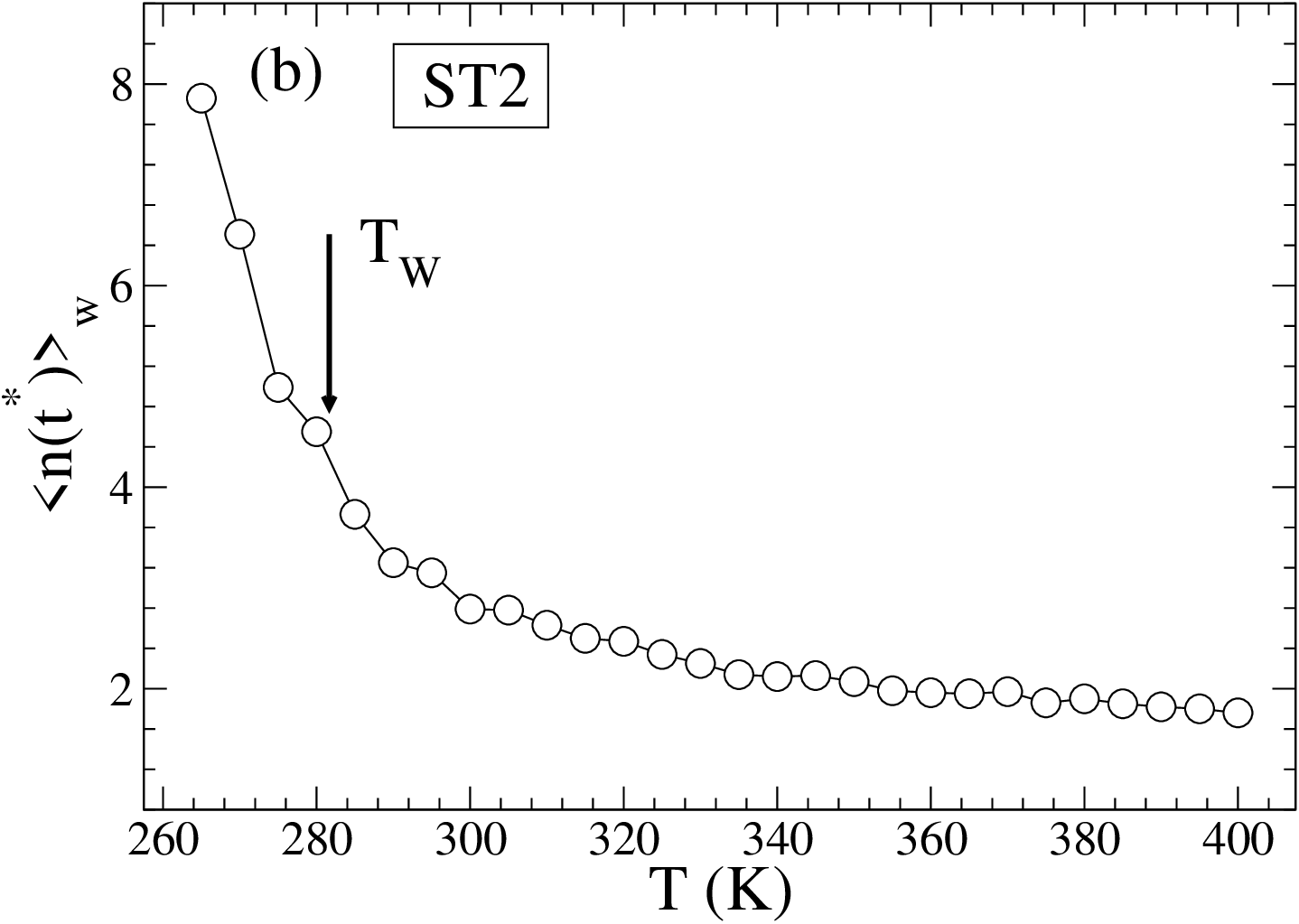}
\end{center}
\caption{Dynamical heterogeneities in the ST2 model of water. (a)
 $\langle n(t)\rangle_w$ as a function of $t$ at 5~K intervals for
 $\rho=0.83$~g/cm$^3$ from $265$~K to $300$~K.  The bold line shows
 the $T=280$~K isotherm where the constant volume specific heat $C_V$
 has a maximum. (b) $\langle n(t^*)\rangle_w$ as a function of $T$
 for $\rho=0.83$~g/cm$^3$.  We indicate the temperature at which
 $C_V$ has a maximum by a vertical arrow.  The maximum of $C_V$
 occurs in the same region where other response functions, such as
 $C_P$, have maxima.}
%$\langle n(t^*)\rangle$  increases markedly near $T<T_W(P)$.}
\label{fig:clus-ST2}
\end{figure}

To further test that the breakdown of the SE relation in water is
associated with the onset of large DH near the Widom line, we show
$\langle n(t)\rangle_w$ for different $T$ along an isochore of density
$\rho=0.83$~g/cm$^3$ for the ST2 model of water in
Fig.~\ref{fig:clus-ST2}(a).  We show an isochore because only
isochoric data are available from Ref.~\cite{poole2}. As in the case
of TIP5P, we find the emergence of a second time scale larger than
$0.5$~ps in $\langle n(t)\rangle_w$ near the crossing of the Widom
line at this density.  Similarly, $\langle n(t^*)\rangle_w$ increases
sharply near the Widom line temperature [see
Fig.~\ref{fig:clus-ST2}(b)]. Hence the sharp growth of DH and the
appearance of a second time scale in $\langle n(t)\rangle_w$ both
occur near the Widom line.  We also find that the magnitude of
$\langle n(t^*)\rangle_w$ is larger for the ST2 model than for the
TIP5P model at the Widom temperature, consistent with the above
observation that the breakdown of the SE relation is more pronounced
for the ST2 model than for the TIP5P model.

\begin{figure}
\begin{center}
\includegraphics[width=10cm]{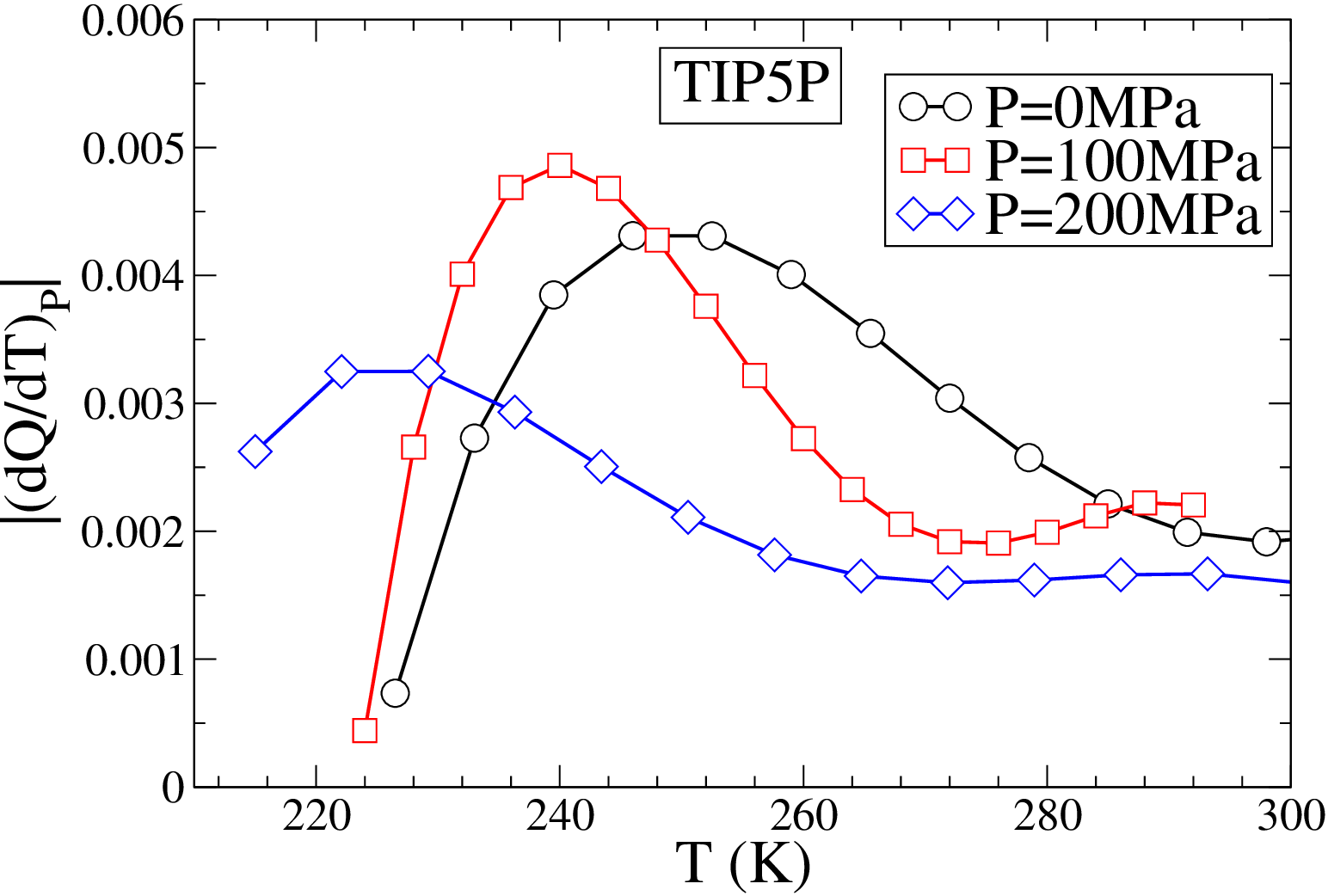}
\end{center}
\caption{ $|(\partial Q/\partial T)_P|$ as a function of $T$ for
$P=0$~MPa and $100$~MPa and $200$~MPa. We note that while the
temperature of the maximum decreases with pressure, the value
$|(\partial Q/\partial T)_P|$ at the maximum for $P=200$~MPa than the
lower pressures. This could be an artifact of the TIP5P potential
which is known not to correctly reproduce the radial distribution
function at high pressures.}
\label{fig:tetrahedrality}
\end{figure}

Finally, we consider the correspondence between DH and static
structural heterogeneity in the supercritical region; this region is
characterized by large fluctuations spanning a wide range of
structures, from HDL-like to LDL-like. To quantify these structural
fluctuations, we calculate for the TIP5P model the temperature
derivative of a local tetrahedral order parameter
$Q$~\cite{erringtonQ}. In Fig.~\ref{fig:tetrahedrality}(a), we show
$|(\partial Q/\partial T)_P|$ at different $T$ for $P=0$~MPa, and
$100$~MPa, and find maxima~\cite{kumarProtein} at
$T_W$(P)~\cite{xuPNAS,Oguni}. The maxima in $|(\partial Q/\partial T)_P|$
indicates that the change in local tetrahedrality is maximal at
$T_W(P)$, which should occur when the structural fluctuation of
LDL-like and HDL-like components is largest. We see that the growth of
the dynamic heterogeneity coincides with the maximum in fluctuation of
the local environment. Also, since $Q$ quantifies the orientational
order, the fact that we find that $|(\partial Q/\partial T)_P|$ has
maximum at approximately the same temperature where $C_P=T(\partial
S/\partial T)_P$ has a maximum, supports the idea that water anomalies
are closely related to the orientational order present in water.

Before concluding, we note that more than a dozen other
phenomena have been correlated with $T_W(P)$.  Some
of these phenomena are from simulations, such as the crossover in the
relaxation time of the fluctuations in orientational order parameter $Q$~\cite{kbs07} or the maximum in the temperature derivative of
the number of hydrogen bonds per molecule ~\cite{kfs06}.  Others are
from only experiments, such as the sharp drop in the the temperature
derivative of the zero-frequency structure factor observed by QENS or
the appearance of a Boson peak, both observed by quasi-elastic neutron
scattering ~\cite{Liu04}.  Finally, some anomalies that correlate with
the Widom line are found in both simulations and experiment,
such as the dynamic ``fragile-to-strong'' crossover in the diffusion
constant, or the sharp drop in temperature derivative of of the mean
squared displacement (seen in QENS, NMR, and simulations)
~\cite{mallamaceJCP,xuPNAS,kumarProtein}.

%Before concluding, we note that our results do not prove that the
%breakdown of SE relation is caused by the rapid restructuring of water
%near the Widom line. In fact, what we see is only the correlation of
%the temperature of the SE relation breakdown and $T_W(P)$. However
%more than a dozen other phenomena have been interpreted as arising
%from crossing $T_W(P)$.  Some of these phenomena are from simulations,
%such as the crossover in the relaxation time of the fluctuations in
%orientational order parameter $Q$~\cite{kbs07} or the maximum in the
%temperature derivative of the number of hydrogen bonds per molecule
%~\cite{kfs06}.  Others are from only experiments, such as the sharp
%drop in the temperature derivative of the zero-frequency structure
%factor observed by QENS or the appearance of a Boson peak, both
%observed by quasi-elastic neutron scattering ~\cite{Liu04}.  Finally,
%some anomalies apparently correlated with crossing the Widom line are
%found in both simulations and experiment, such as the dynamic
%"fragile-to-strong" crossover in the diffusion constant, or the sharp
%drop in temperature derivative of the mean squared displacement (seen
%in QENS, NMR, and simulations)
%~\cite{mallamaceJCP,xuPNAS,kumarProtein}.

In conclusion, we find that the breakdown of the Stokes-Einstein
relation for $P<P_C$ can be correlated with the Widom line. In
particular, rescaling $T$ by $T-T_W(P)$ yields good data collapse of
$D\tau_{\alpha}/T$ for different pressures. Rapid structural changes
occur at $T$ near the Widom line, where larger LDL ``patches'' emerge
upon cooling. The size of the dynamic heterogeneities also increases
sharply as the Widom line is crossed. The breakdown of the SE relation
can be understood by the fact that diffusion at low $T$ is dominated
by regions of fastest moving molecules (DH) while the relaxation of
the system as a whole is dominated by slowest moving molecules.
Consistent with this, we found that the growth of mobile particle
clusters occurs near the Widom line and also near the breakdown of the
SE relation for $P<P_C$. Thus the SE breakdown in water is consistent
with the LL-critical point
hypothesis~\cite{mishima98,poole1,angell-review,pgdbook,pabloreview}.
Our results are also consistent with recent experimental findings in
confined water~\cite{Liu04,mallamaceJCP,chenmallamace2006}.

\medskip

\noindent We thank C. A. Angell, S.-H. Chen, G. Franzese,
J. M. H. Levelt Sengers, S. Han, L. Liu, M. G. Mazza, F. Sciortino,
M. Sperl, K. Stokely, B. Widom, L. Xu, Z. Yan, E. Zaccarelli and
especially S. Sastry for helpful discussions and the NSF Chemistry Program
for support. We also thank the Boston University Computation Center
and Yeshiva University for allocation of computational time.

\newpage

\begin{thebibliography}{99}
\bibitem{footnote} The experimental data for $D$ were taken from
Ref.~\cite{prielmeierXX} and the experimental data for $\eta$ were
taken from Ref.~\cite{angell-review81}. The $T_W$ for experimental
water at $P=1$~atm is taken to be $225$~K~\cite{Oguni} which is also
supported by simulation studies of the SPC/E model of
water~\cite{sas}.
\end{thebibliography}
\end{document}